\documentclass{article}
\usepackage{oldgerm,amsmath,pifont,amsfonts,amssymb,latexsym,bbm}
\usepackage{float,xspace,calc,theorem,ifthen}
\usepackage{enumerate,psboxit,subfigure}
\usepackage[dvips]{graphicx}
\usepackage{hyperref}
\usepackage{cite}
\usepackage{changebar}
\usepackage{verbatim}
\usepackage{color}
\definecolor{greencolor}{rgb}{0,0.5,0.2}
\definecolor{redcolor}{rgb}{.7,0.,0.}
\definecolor{bluecolor}{rgb}{0,0.,1.}
\definecolor{greycolor}{rgb}{.5,.5,.5}

\def\1{\mathbbm{1}}

\def\N{\mathbb{N}}

\newcommand{\rz}{{\mathbb R}}

\begin{document}

\newcommand{\myfont}{\sf }

\title{Escape rate scaling in infinite measure preserving systems}
\author{Sara Munday\footnote{sara.munday@unibo.it}  and Georgie Knight\footnote{georgie.knight@bristol.ac.uk}}
\date{}
\maketitle
\begin{abstract}
We investigate the scaling of the escape rate from piecewise-linear dynamical systems displaying intermittency due to the presence of an indifferent fixed-point. Strong intermittent behaviour in the dynamics can result in the system preserving an infinite measure. We  define a neighbourhood of the indifferent fixed point to be a hole through which points escape and investigate the scaling of the rate of this escape as the length of the hole decreases, both in the finite measure preserving case and infinite measure preserving case. In the infinite measure preserving systems we observe logarithmic corrections to and polynomial scaling of the escape rate with hole length. Finally we conjecture a relationship between the wandering rate and the observed scaling of the escape rate.\\
\\PACS numbers: 05.45.Ac, 05.60.Cd
\end{abstract}

\section{Introduction}
\label{Sec:Intro}

We investigate discrete time dynamical systems that have a hole $H$ in their phase space $X$. In other words,  the iterates of points $x \in X$ are considered only up until they enter $H$, after which they leave the system. In this setting an object of particular interest is the survival probability $Pr(n)$ which is the measure $\mu(S_n)$ of the set of surviving points $S_n$ up to time $n$,  and $\mu$ is the measure of initial conditions in the system. When the dynamics are strongly chaotic, that is, when they exhibit exponential decay of correlations, we expect the survival probability to decay exponentially. The rate of this decay is defined to be
\begin{equation}
	               \gamma:= \lim_{n \rightarrow \infty} -\frac{\ln(Pr(n))}{n},
\label{Eq:EscapeRateDef}
\end{equation}
where $\gamma$, which is the object of investigation here, is called the {\em escape rate}. While Eq.(\ref{Eq:EscapeRateDef}) tells us that the escape rate is dependent on the initial distribution $\mu$, note that the escape rate $\gamma$ can also be calculated as the largest eigenvalue $\lambda_H$ of the Perron-Frobenius transfer operator of the open system
\begin{equation}
	               \gamma= -\ln(\lambda_H).
\label{Eq:EscapeRateDefEig}
\end{equation}
Therefore, being a spectral quantity, the escape rate is equal for a large class of initial distributions. See \cite{DeYo06} for a discussion.

Systems where iterates can escape through holes were first studied in the context of Markov chains with absorbing states (see \cite{FeEtal95} and references therein) before being discussed in the theory of dynamical systems in \cite{PiYo79}. In this setting initial work was on the study of the conditionally invariant measures of the open system \cite{ColletEtal97}. These are the eigenvectors of the transfer operator of the open system. In particular, one question asked is which conditionally invariant measures are physical, or natural in that they somehow capture the statistics of the open system, see \cite{LivMau03} for more recent work and \cite{DeYo06} for a review. In addition to the mathematical interest, there is much attention given to the escape rate and its direct applications by the physics community. See \cite{AltRev13,DetChapter} for recent reviews.

A topic which has recently received a lot of interest has been the study of the escape rate as a function of certain system parameters, for instance studying the behaviour of the escape rate as a function of hole position, or size. Initial numerical work in \cite{PaPa97} uncovered a nontrivial relationship between the periodic orbit structure and the lifetime of transients in chaotic maps. Related rigorous results from \cite{KeLi09} show that the asymptotic behaviour of the escape rate as the hole shrinks to a point is dependent upon the stability of the limiting point. This is independently shown in \cite{BuYu11} for the doubling map; they also show that finite size holes of equal measure can have different escape rates, among other results. Further work on the asymptotic behaviour of the escape rate can be found in \cite{Cr13} and \cite{Det13} where higher order logarithmic corrections are found. In \cite{DeWr12} it is shown that for a large class of systems the escape rate as a  function of the continuous varying of the hole through phase space is a devil's staircase.

Much work has been done on the uniformly hyperbolic setting. However, intermittency in open dynamical systems can lead to some interesting properties. Systems which display intermittency but still preserve finite measures are studied in \cite{Dahl99} \cite{FrMuSt11} \cite{DeFe14}, where it is shown that the intermittency can lead to polynomial rather than exponential rates of escape and the non-existence of a natural conditionally invariant measure, which means that the escape rate depends heavily on the initial set-up of the system. Here we intend to study deviations from the standard linear scaling (see for example \cite{FrMuSt11}) of the escape rate with hole length. It is shown in \cite{KeLi09} that for a particular example, the escape rate does not scale linearly with the length of the hole when the measure of the closed system vanishes at the limiting point of the hole. Here our main goal is to examine and understand the scaling when the underlying closed map preserves an infinite invariant measure and when the measure diverges at the limit point.  Rather than approach this problem with the elegant perturbation theory used in \cite{KeLi09}, where the small hole is considered to be a perturbation of the dynamics of the closed system, we approach the problem using the {\em top down} techniques from \cite{Cr13} where one starts with a large hole and shrinks to the closed system.

The paper is organised as follows. In Section \ref{Sec:Prelim} we define the setting of a family of piecewise-linear, $\alpha$-Farey maps, where $\alpha$ is an infinite partition of the unit interval. Depending on the type of partition, the $\alpha$-Farey map preserves either an infinite or finite Lebesgue-absolutely continuous measure. In Section \ref{Sec:Derive-zeta} we then study the escape rate in these systems with a hole by deriving the dynamical zeta function, a polynomial which encodes important properties of the dynamics. Via the zero of this polynomial, in Section \ref{Sec:Asymp} we study the small hole asymptotic behaviour of the escape rate and derive how the escape rate scales with the length of the hole for some particular examples of $\alpha$-Farey maps and for a parameter-dependent version which incorporates both finite and infinite measure preserving dynamics as a function of the control parameter. Finally in Section \ref{Sec:Conclusion} we make some concluding remarks and observe that the scaling found in this particular setting is intimately related to a quantity known as the wandering rate, we furthermore conjecture that this relationship will hold in more general systems.

\section{Preliminaries}
\label{Sec:Prelim}
Let $\alpha:=\{A_n:n\in\N\}$ denote a countably infinite partition of the interval $(0, 1]$, consisting of non-empty, right-closed and left-open intervals, and let $a_n:=\varepsilon(A_n)$, where $\varepsilon$ denotes the Lebesgue measure,  and
\begin{equation}
                        t_n:=\sum_{k=n}^\infty a_k.
\label{Eq:tn}
\end{equation}
 It is assumed throughout that the elements of $\alpha$ are ordered from right to left, starting from $A_1$, and that these elements accumulate only at the origin. Let us denote the set of all such partitions by $\mathcal{A}$. Then, for a given partition $\alpha\in \mathcal{A}$, the  map $F_{\alpha}:[0,1] \to [0,1]$ is given by
\begin{equation}
F_{\alpha}(x):=\left\{
        \begin{array}{ll}
          (1-x)/a_1 & \hbox{if $x\in A_1$,} \\
          {a_{n-1}}(x-t_{n+1})/a_{n}+t_n & \hbox{if $x\in A_n$, for  $n\geq2$.}
\\
	  0 & \hbox{if $x=0$. }
        \end{array}
      \right.
\label{Eq:A-Farey-map}
\end{equation}
For example, let us consider the dyadic partition
$\alpha_D:=\left\{\left(1/2^{n},1/2^{n-1}\right]:n\in\N\right\}$. One can  immediately verify that
the map  $F_{\alpha_D}$ coincides with  the tent map $T:[0,1]\to[0,1]$, which is given by
\begin{equation}
T(x):=\left\{
                      \begin{array}{ll}
                                                                                                     2x, & \hbox{for $x\in[0,1/2)$;} \\
                                                                                                     2-2x, & \hbox{for $x\in [1/2, 1]$.}
                                                                                                   \end{array}
                                                                                                 \right.
\label{Eq:Tentmap}
\end{equation}
To see this,  it is enough to note that for each $n\in\N$ we have that $a_n=2^{-n}$ and $t_n=2^{-(n-1)}$.

\begin{figure}[ht!]
\begin{center}
\includegraphics[width=0.33\textwidth]{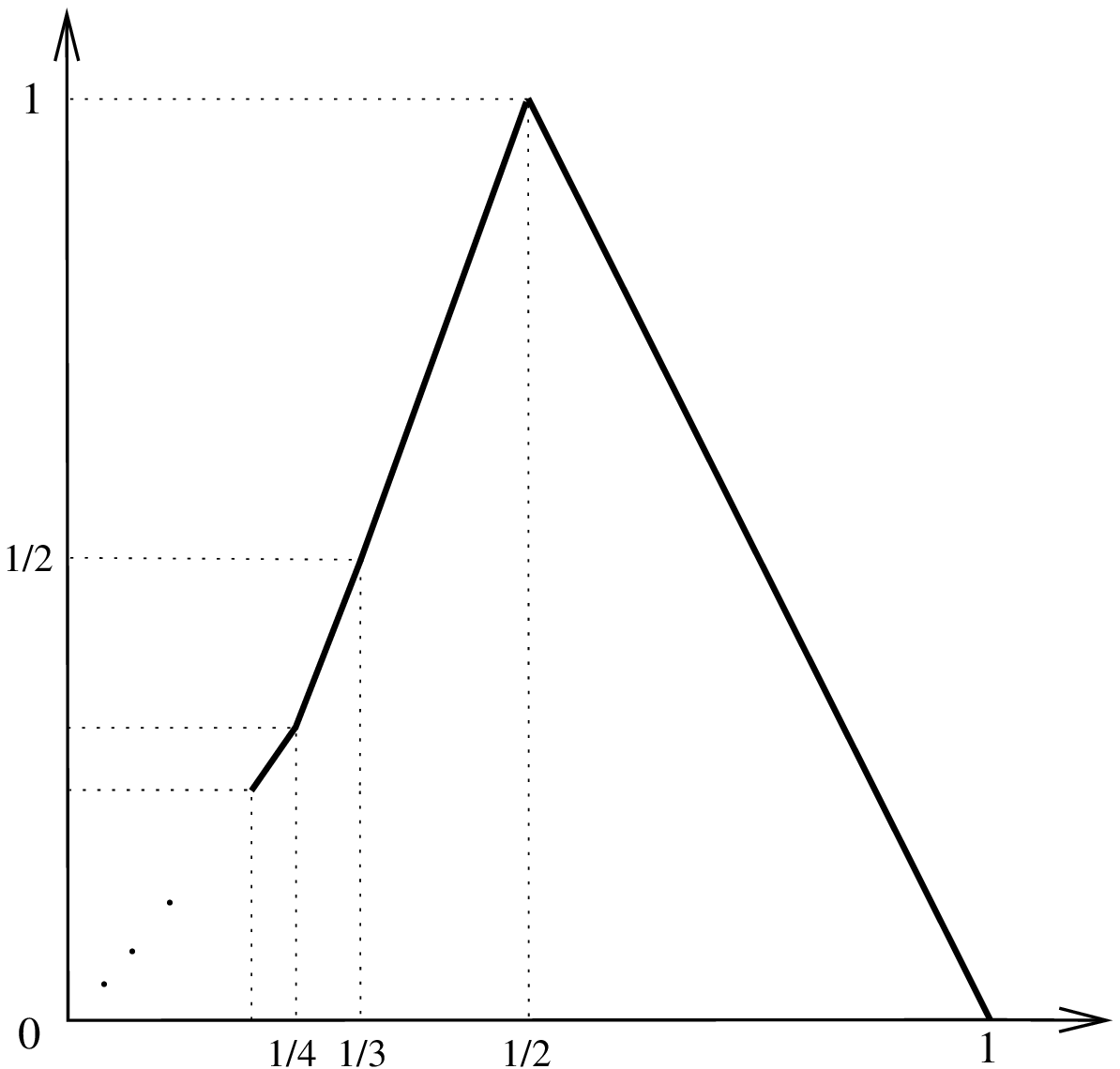}\hspace{0.1\textwidth}
\includegraphics[width=0.33\textwidth]{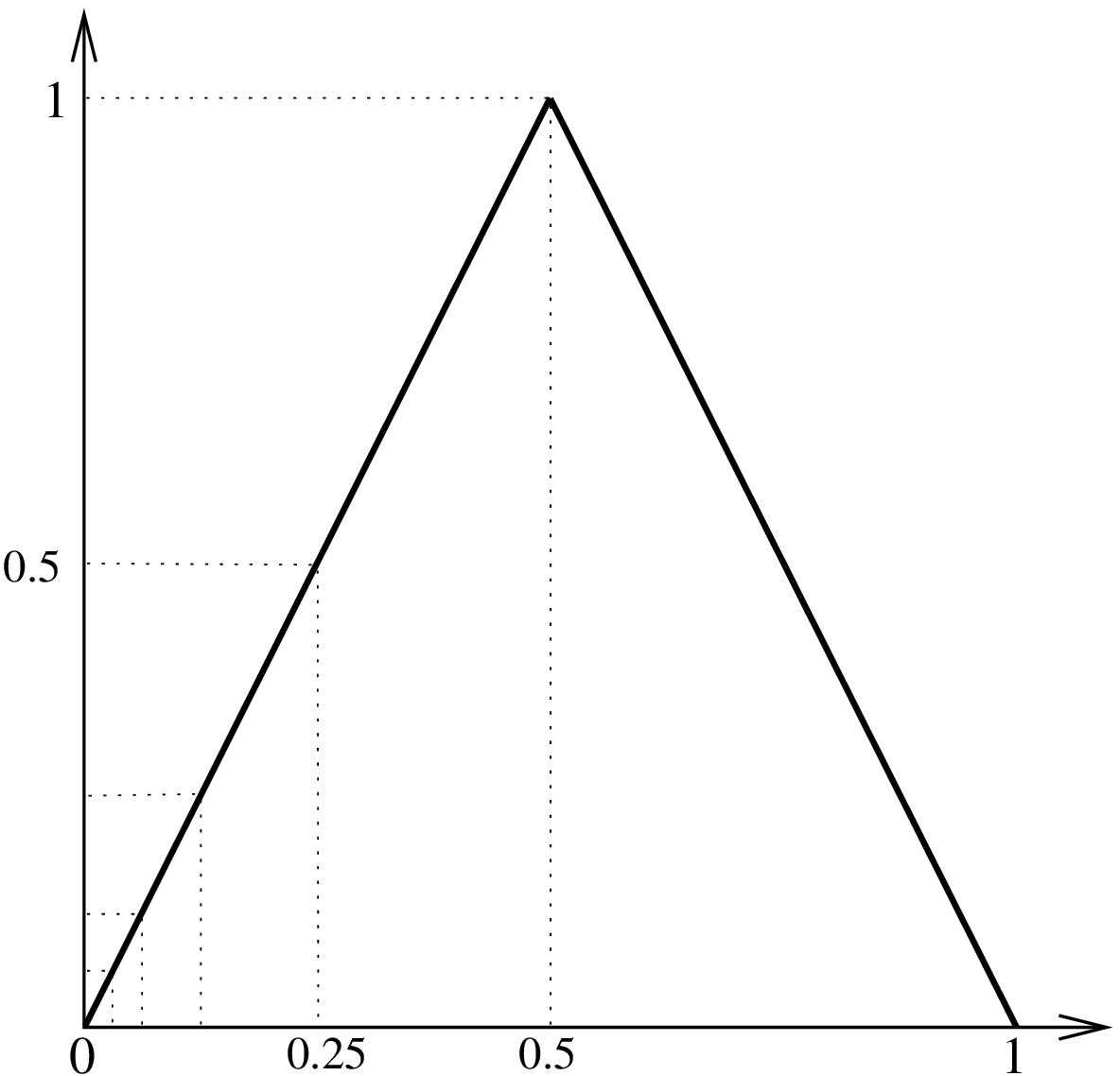}
\caption{The $\alpha_H$-Farey map, where $t_{n}=1/n$, $n\in \N$, and the $\alpha_D$-Farey map, where $t_n=2^{-(n-1)}$.}\label{Fig:alphafarey}\end{center}
\end{figure}

Alternatively consider the partition $\alpha_H:=\left\{A_n:=\left(1/(n+1),1/n\right]:n\in\N\right\}$, which we will call the {\em harmonic partition}. Here, we obtain the map $F_{\alpha_H}$ which is given explicitly by
\begin{equation}
F_{\alpha_H}(x):=\left\{
        \begin{array}{ll}
          2-2x & \hbox{for $x\in (1/2, 1]$;} \\
          \frac{n+1}{n-1}x-\frac1{n(n-1)} & \hbox{for $x\in (1/(n+1),1/n]$ .}
        \end{array}
      \right.
\label{Eq:Farey-map}
\end{equation}
The graphs of the maps $F_{\alpha_D}$ and $F_{\alpha_H}$ are shown in figure \ref{Fig:alphafarey}. For more details about the $\alpha$-Farey systems, we refer the reader to \cite{KMS}.

It was shown in \cite{KMS} that the Lebesgue-absolutely continuous  measure $\nu_\alpha$ defined by the density $\phi_{\alpha}$ which is given  by
\begin{equation}
\phi_{\alpha}:=\frac{d\nu_\alpha}{d\varepsilon}=\sum_{n=1}^{\infty} \frac{t_n}{a_n}  \cdot \mathbbm{1}_{A_{n}},
\end{equation}
is the unique, up to multiplication by a constant,  Lebesgue-absolutely continuous measure (ACIM) preserved by $F_\alpha$. Furthermore, this measure is infinite (non-normalisable) if and only if
\begin{equation}
\lim_{n \rightarrow \infty}\sum_{k=1}^n t_k = \infty.
\label{Eq:Infinite-measure-property}
\end{equation}
For the two examples mentioned above, we have that for $F_{\alpha_D}$, the density is the function constantly equal to 2 (in keeping with the fact that we know already in this case that the invariant measure is simply the Lebesgue measure), and for $F_{\alpha_H}$, the density is the simple function given by $\phi_{\alpha_H}= \sum_{n=1}^{\infty} (n+1)\cdot \mathbbm{1}_{A_n}$.

We open the system $([0, 1], F_\alpha)$, for an arbitrary $\alpha\in \mathcal{A}$ by defining the interval  $[0, t_n]$ to be the hole.
The length of the hole will take on a discrete set of value s, $\varepsilon([0,t_n])$, depending on $n$,  which tend to zero as $n$ goes to infinity (we will write $\varepsilon$ in the following to refer to the length of the hole). Here, the Perron-Frobenius transfer operator of the closed system $P$ acts on functions that are piecewise constant on the intervals $A_n$. As mentioned already above, there exists a unique ACIM $\nu_\alpha$ of the closed system, with corresponding eigenvalue $1$, that is $P \nu_\alpha= \nu_\alpha$.

The corresponding transfer operator of the open system $P_{[0,1]\backslash H}$ acts on functions that are piecewise constant on the intervals $A_1,..,A_k$ and the interval $[0,t_n]$. In this setting of sub-shift of finite type with a cylindrical hole, we have that the transfer operator $P_{[0,1]\backslash H}$ has an eigenfunction correponding to an absolutely continuous with respect to Lebesgue, conditionally invariant measure (ACCIM) $\varphi_H$ and corresponding eigenvalue $\lambda_H$. That is $P_{[0,1]\backslash H} \varphi_H=\lambda_H \varphi_H$ \cite{Collet00}. The behaviour of $\lambda_H$, and hence the escape rate, as a function of hole size is the subject of this paper. 
\section{Deriving the dynamical zeta function of the open system}
\label{Sec:Derive-zeta}
The Perron-Frobenius transfer operator $P_{[0,1]\backslash H}$ in the setting we consider, for the hole $[0, t_n]$, takes the form of a transfer matrix with the  the following structure
\begin{equation}
 \mathbf{P}_n=\left( \begin{array}{cccccc}
0        &0&0&0&0          &t_n \\
0        &0&0&0&0          & a_{n-1} \\
0        &1&0&0&0          & a_{n-2} \\
0        &0&1&0&0          & a_{n-3} \\
\vdots& &  &\ddots&\vdots & \vdots \\
0        &0&0&0&1          &a_1 \\
\end{array} \right)
\label{Eq:P}
\end{equation}
Note that the eigenvalue $\lambda_H$ is equal to the inverse of the zero $z_0$ of the polynomial $\det(1-z\mathbf{P}_n)$. This polynomial is also known as the dynamical zeta function, $\zeta_{op}(z):=\zeta_{op, n}(z)$, where the subscript {\em op} refers to the open system and, to lighten the notation, we drop the 
dependence on $n$. Let us now derive an expression for $\zeta_{op}(z)$. We have that
\begin{equation}
 \zeta_{op}(z)=\left| \begin{array}{cccccc}
1        &0&0&0&0          &-zt_n \\
0        &1&0&0&0          & -za_{n-1} \\
0        &-z&1&0&0          & -za_{n-2} \\
0        &0&-z&1&0          & -za_{n-3} \\
\vdots& & &\ddots&\vdots & \vdots \\
0        &0&0&0&-z          &1-za_1 \\
\end{array} \right|
\label{Eq:1-zp}
\end{equation}
Multiplying the second row by z and adding to the third, and continuing this process to the $n^{th}$ row we obtain
\begin{equation}
\zeta_{op}(z)=\left| \begin{array}{cccccc}
1        &0&0&0&0          &-zt_n \\
0        &1&0&0&0          & -za_{n-1} \\
0        &0&1&0&0          & -za_{n-2}-z^2 a_{n-1} \\
0        &0&0&1&0          & -za_{n-3} -z^2a_{n-2}-z^3a_{n-1} \\
\vdots& & &\ddots&\vdots & \vdots \\
0        &0&0&0&0         &1-\sum_{k=1}^{n-1}z^ka_k \\
\end{array} \right|
\label{Eq:1-zp2}
\end{equation}
Expanding the determinant along the first column and then the last row we obtain the dynamical zeta function
\begin{equation}
                       \zeta_{op}(z)=\det(1-z\mathbf{P}_n)= 1-\sum_{k=1}^{n-1}z^ka_k.
\label{Eq:zeta}
\end{equation}
\section{Asymptotic behaviour and scaling}
\label{Sec:Asymp}
We remind the reader that we will use $\varepsilon$ to refer to the length of the hole $\varepsilon:=\varepsilon([0,t_n])$. In \cite{KeLi09} it was shown that for a large class of hyperbolic maps
\begin{equation}
                           \gamma=\varepsilon(1-C)+o(\varepsilon)
\label{Eq:KLResult}
\end{equation}
where $C$ is a constant that depends upon the stability eigenvalue of the limiting point of the hole. We recall that the notation $g(\varepsilon)=o(f(\varepsilon))$ means that $\lim_{\varepsilon \rightarrow 0} g(\varepsilon)/f(\varepsilon)=0$. For example in the doubling map, Eq.(\ref{Eq:KLResult}) translates as
\begin{equation}
                           \gamma=\varepsilon(1-2^{-p})+o(\varepsilon)
\label{Eq:BSResult}
\end{equation}
where $p$ is the period of the limiting point. This first order result states that for small holes the escape rate behaves linearly in $\varepsilon$. For results on higher order terms see \cite{Det13}, \cite{Cr13}. We will investigate deviations from this linear asymptotic behaviour in the family of $\alpha$-Farey maps defined above in Eq.(\ref{Eq:A-Farey-map}) using the dynamical zeta function derived in Eq.(\ref{Eq:zeta}). To this end, we aim to find a function $f:[0,1]\rightarrow [0,\infty)$ with the properties that $\lim_{\varepsilon \rightarrow 0}f(\varepsilon)=0$ and
\begin{equation}
                       \gamma=f(\varepsilon)+o(\varepsilon).
\label{Eq:Escape_f}
\end{equation}
That is, the function $f$ gives the small-hole behaviour of the escape rate. Given the results described by Eq.(\ref{Eq:KLResult}) and Eq.(\ref{Eq:BSResult}), in the case of hyperbolic maps $f$ will be a linear function of $\varepsilon$. Using the relationship between the zero $z_0$ of the polynomial in Eq.(\ref{Eq:zeta}) and the escape rate, namely,  $\gamma=\ln(z_0)$, we can write the zero $z_0$ in terms of the function $f$
\begin{equation}
                         z_0=1+f(\varepsilon) + o(f(\varepsilon)).
\label{Eq:Zeta-zero}
\end{equation}
The explicit form of the function $f(\varepsilon)$ remains to be derived and will depend on the map $F_{\alpha}$. We will use our expression for the dynamical zeta function Eq.(\ref{Eq:zeta}) and Eq.(\ref{Eq:Zeta-zero}) to do this. Using Eq.(\ref{Eq:Zeta-zero}) in Eq.(\ref{Eq:zeta}) gives,
\begin{equation}
                           \zeta_{op}(z_0)=1-\sum_{k=1}^{n-1}(1+f(\varepsilon)+o(f(\varepsilon))^ka_k=0.
\label{Eq:Zeta_with_zero}
\end{equation}
Solving Eq. (\ref{Eq:Zeta_with_zero}) and letting $\varepsilon$ tend to 0 (or, in other words, letting $n\to\infty$), we will be able to determine the 
function $f(\varepsilon)$, where we remind the reader that this function depends upon the partition $\alpha$. Using,
\begin{equation}
                        \left(1+f(\varepsilon)+o(f(\varepsilon)\right)^k=\sum_{j=0}^k {k \choose j} (f(\varepsilon))+o(f(\varepsilon))^j,
\label{Eq:expansion_polynomial}
\end{equation}
rearranging the order of the summations and extracting the $j=0$ term we obtain,
\begin{equation}
                         \zeta_{op}(z_0) = 1-\sum_{k=1}^{n-1}a_k-\sum_{j=1}^{n-1}\sum_{k=j}^{n-1}{k \choose j}(f(\varepsilon))+ o(f(\varepsilon))^ja_k=0. \label{Eq:Zeta-approx}
\end{equation}
Note that from the definition of the partition $\alpha$ we have that
\begin{equation}
\sum_{k=1}^{\infty}a_k=1.
\label{Eq:ak_full_sum}
\end{equation}
From the definition of $t_n$ given in Eq.(\ref{Eq:tn}) and of the length of the hole $\varepsilon$ it immediately follows that
\begin{equation}
                           \sum_{k=1}^{n-1}a_k=1-\varepsilon.
\label{Eq:aksum}
\end{equation}
Using Eq.(\ref{Eq:aksum}) in  Eq.(\ref{Eq:Zeta-approx}) we therefore obtain
\begin{equation}
                          \varepsilon-\sum_{j=1}^{n-1}\sum_{k=j}^{n-1}{k \choose j}(f(\varepsilon))+ o(f(\varepsilon))^ja_k=0.
\label{Eq:s1Sum}
\end{equation}
Also, we note here one further relation which will be useful in what follows:
\begin{equation}
                           \sum_{k=j}^{n-1}{k \choose j} a_k= \sum_{k=j}^{n-1}{k \choose j}(t_k-t_{k+1})= -{n \choose j}t_n+ \sum_{k=j}^{n}{k-1 \choose j-1}t_k.
\label{Eq:kakSum}
\end{equation}
Let us now consider some particular examples, including cases where the invariant measure of the closed system is finite and where it is infinite, in order to illustrate the usefulness of Eqs. (\ref{Eq:zeta}) and (\ref{Eq:s1Sum}).

\subsection{Tent map}
\label{SubSec:tentmap}
Consider the dyadic partition, that is, the partition defined by $t_k=2^{-(k-1)}$ as described above. For each $n$ we have the system open on the interval $[0,2^{-(n-1)}]$ giving $\varepsilon=2^{-(n-1)}$, implying that $n=\ln(1/\varepsilon)/\ln(2)+1$.  Using Eq.(\ref{Eq:s1Sum}), Eq.(\ref{Eq:kakSum}) and writing in terms of $n$ for simplicity (as $\varepsilon$ is a function of $n$) we derive,
\begin{equation}
                           2^{-(n-1)}-\sum_{j=1}^{n-1}(f(n)+of(n))^j\left(-{n \choose j}2^{-(n-1)} +\sum_{k=j}^n{k-1 \choose j-1}2^{-(k-1)}\right)=0
\label{Eq:TentmapkakSum}
\end{equation}
{Multiplying Eq.(\ref{Eq:TentmapkakSum}) by $2^{n-1}$  and solving the $j=1$ term in the sum we obtain that}
\begin{eqnarray}\nonumber
                       1- 2^{n-1}(f(n)+of(n))(-n2^{-(n-1)} +2-2^{1-n}) &-&\\
                      2^{(n-1)}\sum_{j=2}^{n-1}(f(n)+of(n))^j\left(-{n \choose j}2^{-(n-1)} +\sum_{k=j}^n{k-1 \choose j-1}2^{-(k-1)}\right)&=&0
\label{Eq:TentmapkakSum2}
\end{eqnarray}
Letting $f(n)=C2^{-(n-1)}$ in Eq.(\ref{Eq:TentmapkakSum2}) with $C$ a constant and letting $n$ tend to infinity shows us that all other terms go to zero whilst $C=0.5$. Converting back into terms of $\varepsilon$ we obtain that $f(\varepsilon)=\varepsilon/2$, as expected \cite{KeLi09}. In other words, the escape rate scales linearly with the length of the hole. We further note that Eq.(\ref{Eq:TentmapkakSum}) implies that there are logarithmic corrections in the higher order terms, a result which is well known (see \cite{Cr13}, \cite{Det13}). See figure \ref{Fig:asymptoticbehaviour} for an illustration of this result.
\subsection{Piecewise Farey map}
\label{SubSec:PFM}
\begin{figure}[tb]
\begin{center}
\includegraphics[width=7cm]{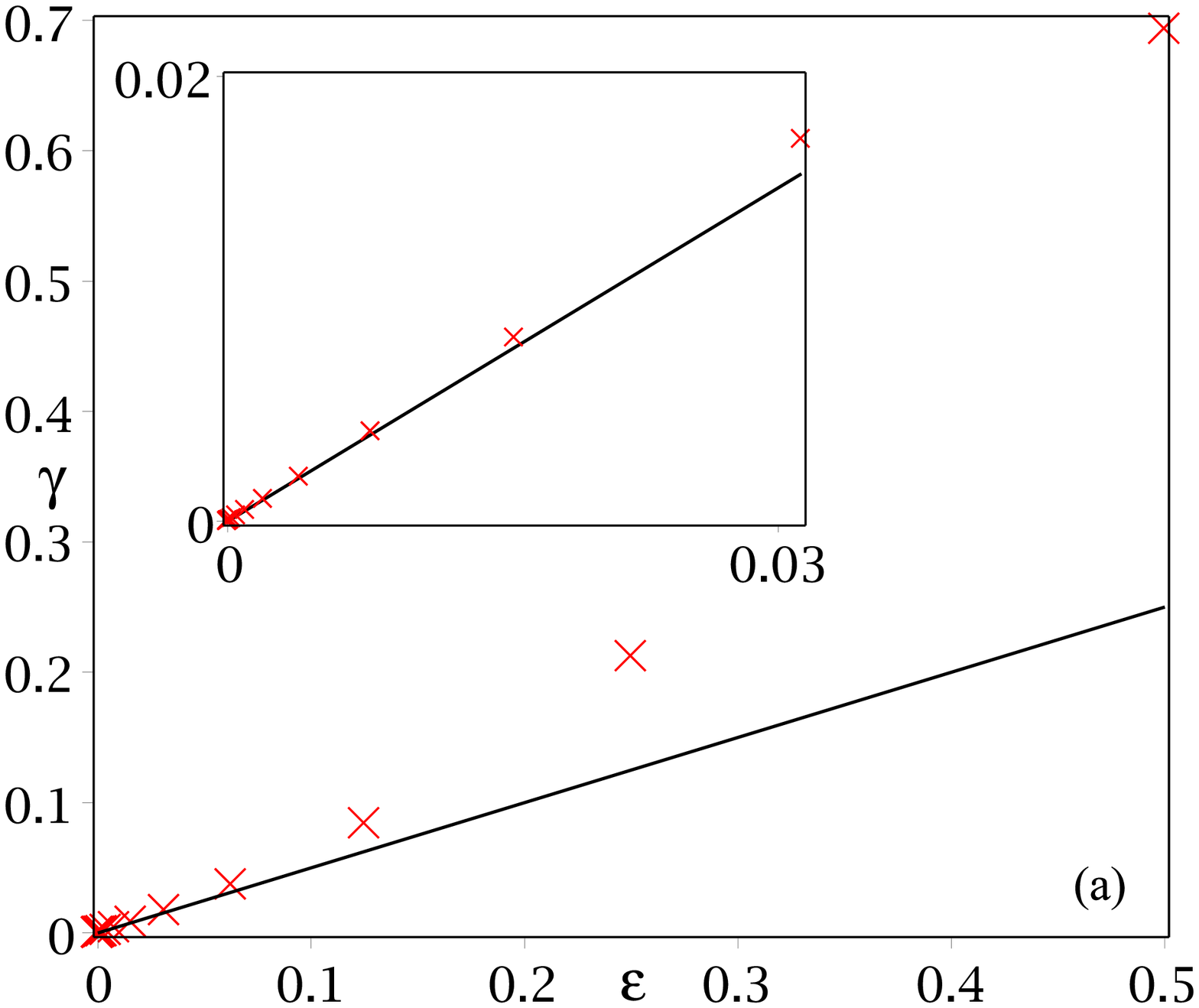} \includegraphics[width=7cm]{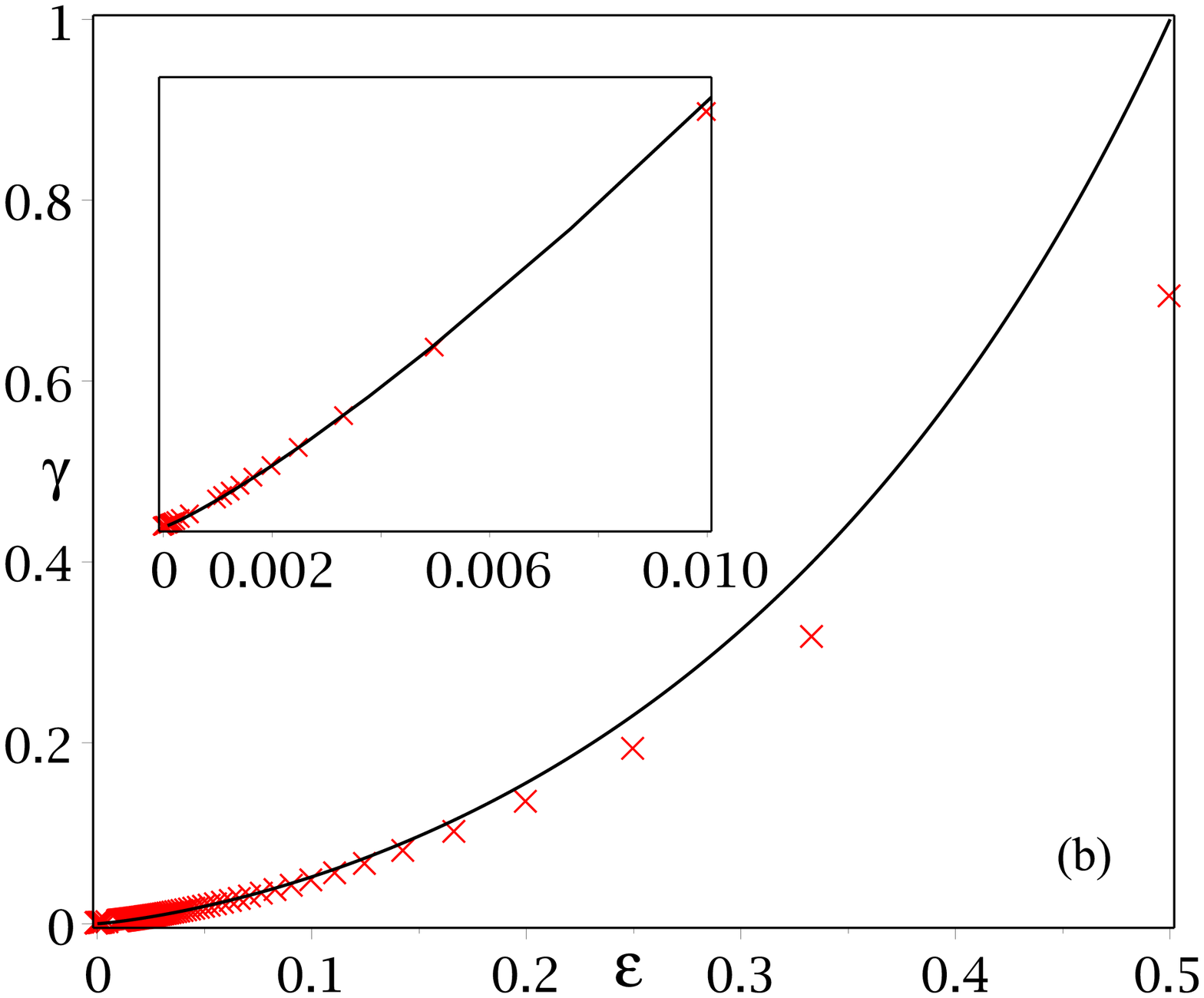}
\caption{Escape rate as a function of hole length. In this figure the escape rate $\gamma$ is plotted as a function of hole size $\varepsilon$ , computed from the smallest zero of the dynamical zeta function (red crosses) along with the approximation $f(\varepsilon)$ (black line) which gives the small hole scaling and asymptotic behaviour. In $\mathbf{(a)}$ these are illustrated for the tent map which admits a finite invariant measure whilst in $\mathbf{(b)}$ the $\alpha_{H}$-Farey map which has an infinite  invariant measure.}\label{Fig:asymptoticbehaviour}\end{center}
\end{figure}

Let us now consider the piecewise linearisation $F_{\alpha_H}$ of the Farey map given by the harmonic partition. As noted above, this map preserves an infinite measure. Here $t_k=1/k$ and $\varepsilon=1/n$.  Again from  Eq.(\ref{Eq:s1Sum}) and Eq.(\ref{Eq:kakSum}) we have in this case
\begin{equation}
                                          \frac{1}{n}-\sum_{j=1}^{n-1}(f(n)+of(n))^j\left(-{n \choose j}\frac{1}{n} +\sum_{k=j}^n{k-1 \choose j-1}\frac{1}{k}\right)=0 \label{Eq:HarmonicSum}
\end{equation}
Multiplying Eq.(\ref{Eq:HarmonicSum}) by $n$ and extracting the $j=1$ term we obtain,
\begin{equation}
                          1-n(f(n)+of(n))(H_n-1)-\sum_{j=2}^{n-1}(f(n)+of(n))^j\left(-{n \choose j} +n\sum_{k=j}^n{k-1 \choose j-1}\frac{1}{k}\right)=0
\label{Eq:HarmonicmapkakSum2}
\end{equation}
Here, $H_n$ is the $n^{th}$ harmonic number which grows logarithmically with $n$, that is,  logarithmically inversely with $\varepsilon$. Where $C$ is a 
constant to be determined, we now let $f(n)=C/n(H_n-1)$ and let $n$ go to infinity in Eq.(\ref{Eq:HarmonicmapkakSum2}). Expanding the binomial coefficient as a polynomial and applying the Euler-Maclaurin formula we see that the higher order summation terms decay to zero as $n$ goes to infinity leaving $C=1$. In terms of $\varepsilon$ we therefore have
\begin{equation}
                          f(\varepsilon)=\frac{\varepsilon}{H_n-1}.
\label{Eq:s1Harmonic}
\end{equation}
However, given that $H_n$ grows logarithmically as $\varepsilon$ goes to zero, we see that the escape rate does not scale linearly with $\varepsilon$, rather there are logarithmic corrections present. See figure \ref{Fig:asymptoticbehaviour}.
\subsection{Parameter dependence}
\label{subsec:pd}
The two examples given above, namely the $\alpha$-Farey maps arising from the dyadic and harmonic partitions, shows the existence of a fundamental difference in the scaling of the escape rate depending on whether the closed system preserves an infinite invariant measure or a finite one. We now explore this further. Let us consider the case that
\begin{equation}
                         t_k=\frac{1}{k^\theta},\ \ k>1,
\label{Eq:t_kParameter}
\end{equation}
where $\theta \in \rz$ is a control parameter of the system. We have $\varepsilon=1/n ^{\theta}$ and from the considerations of Eq.(\ref{Eq:Infinite-measure-property}) for $\theta \leq 1$ the ACIM is infinite whilst for $\theta>1$ it is finite. Note that the sum $\sum_{k=1}^{\infty}  t_k$ is the value of the Riemann zeta function $\zeta(\theta)$ at $\theta$ and that for $\theta=1$ we recover the  $\alpha_{H}$-Farey map. For connections between the Riemann hypothesis and the escape rate see \cite{BuDett05}, \cite{DeRa14}. From  Eq.(\ref{Eq:s1Sum}) and Eq.(\ref{Eq:kakSum})  we have,
\begin{equation}
                                            1-\sum_{j=1}^{n-1}(f(n)+of(n))^j\left(-{n \choose j}+n^{\theta}\sum_{k=j}^n{k-1 \choose j-1}\frac{1}{k^{\theta}}\right)=0
\label{Eq:PArSum}
\end{equation}
\subsection{Finite ACIM}
Using Stieltjes integration, one can write the Riemann zeta function in the following way:
\begin{equation}
\zeta(\theta):=\sum_{k=1}^\infty\frac{1}{k^\theta} = \frac{\theta}{1-\theta} - \theta\int_1^\infty \frac{\{x\}}{x^{\theta+1}}\ dx,
\label{Eq:Zeta-expression}
\end{equation}
where $\{x\}$ denotes the fractional part of $x$. This relation is valid for all $\theta>1$. It therefore follows, after a straightforward calculation, that for $\theta>1$ we have
\begin{equation}
                          \sum_{k=1}^n \frac{1}{k^\theta}=  \zeta(\theta) +\frac{1}{n^{\theta-1}}- \theta\int_{n}^\infty\frac{\lfloor x\rfloor}{x^{\theta+1}}\ dx.
\label{Eq:Sum-finite-case}
\end{equation}
For the integral in Eq.(\ref{Eq:Sum-finite-case}), we have an upper bound given by
\begin{equation}
                          \theta\int_{n}^\infty\frac{\lfloor x\rfloor}{x^{\theta+1}}\ dx <\frac{\theta}{\theta-1}n^{1-\theta}.
\label{Eq:Finite-integral-upperbound}
\end{equation}
Considering the $j=1$ term in Eq.(\ref{Eq:PArSum}), and again with $C$ a constant to be determined,  we let $f(n)=C/n^{\theta}$. We let $n$ go to infinty and as before we expand the binomial coefficient as a polynomial and use the Euler-Maclaurin summation formula to check that the remaining terms is Eq.(\ref{Eq:PArSum}) go to zero. This leaves us with $C=\zeta(\theta)$

\begin{figure}[htb]
\begin{center}
\includegraphics[width=7cm]{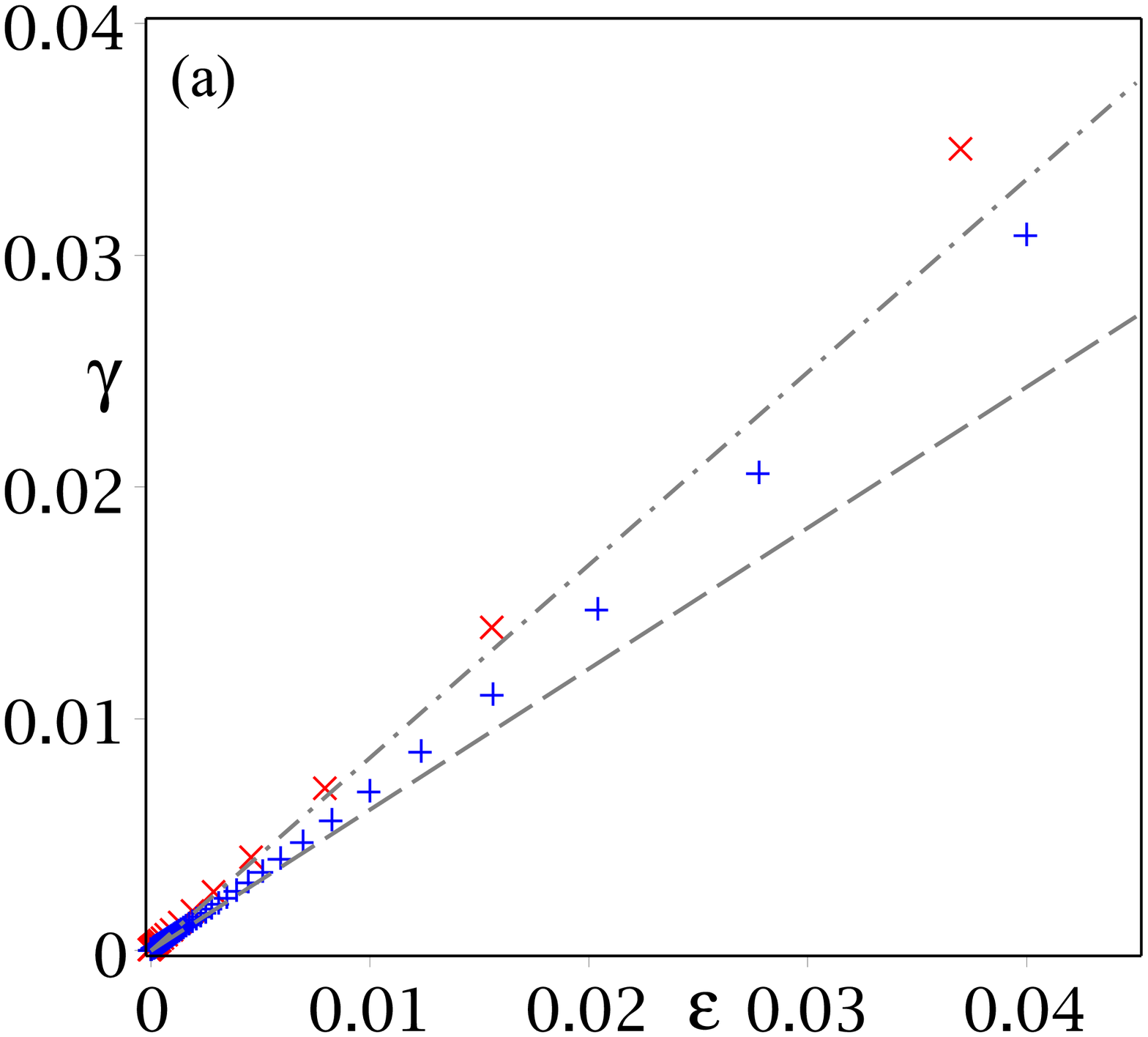} \includegraphics[width=7cm]{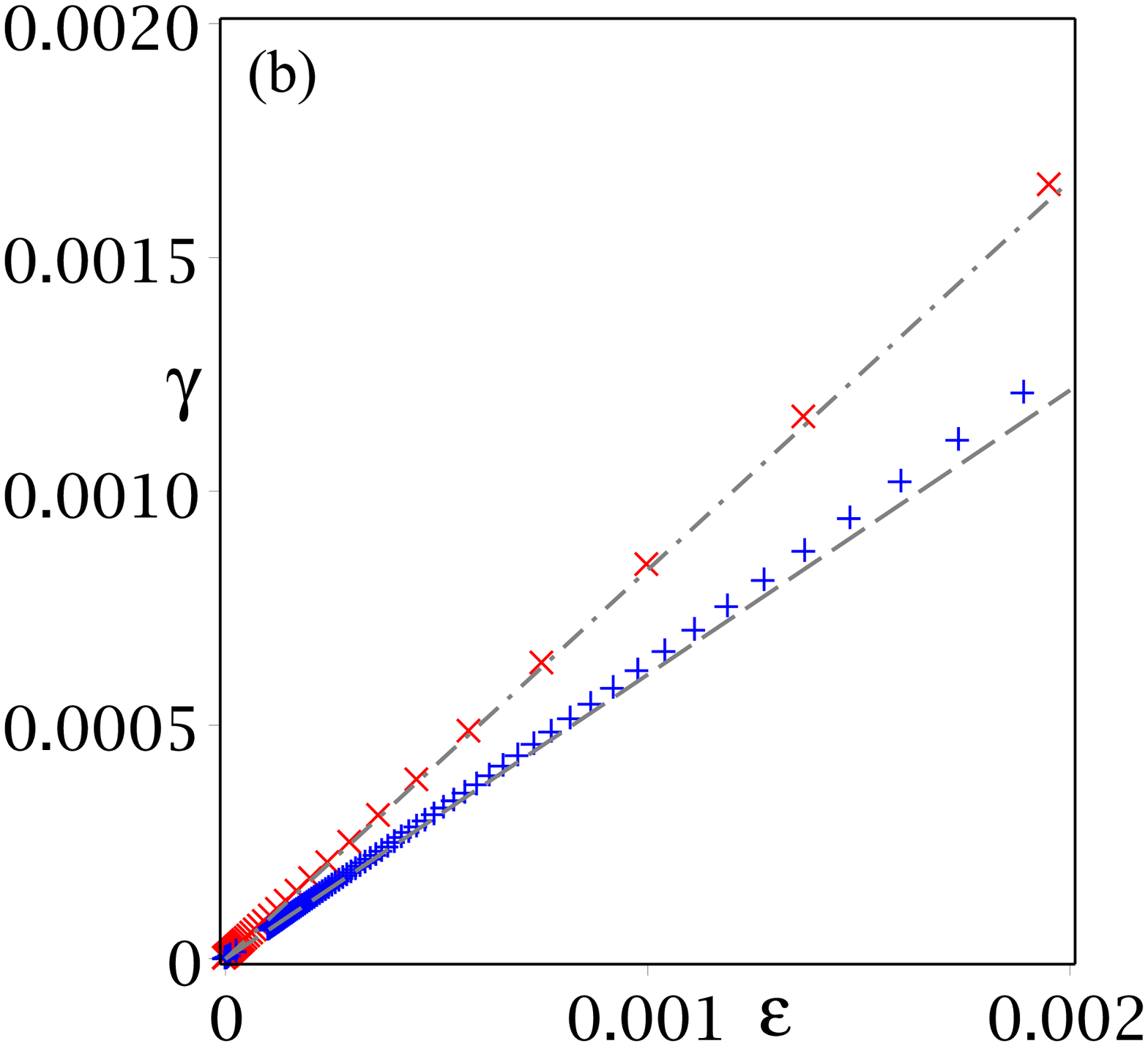}
\caption{Escape rate as a function of hole length. In this figure the escape rate $\gamma$ is plotted as a function of hole size $\varepsilon$ for the $\alpha$ Farey map for parameter value $2$ (red diagonal crosses) and $3$ (blue crosses) along with the small $\varepsilon$ scaling and asymptotic behaviour as given by Eq.(\ref{Eq:s1-finite}) for $\theta=2$  (dot-dash grey line) and $\theta=3$  (dashed grey line). }\label{Fig:asymptotic-behaviour-parameter}\end{center}
\end{figure}

 Hence,   as in the example of the Tent map above, the escape rate scales linearly with $\varepsilon$ with constant $\zeta(\theta)$. That is, $f(\varepsilon)$ is given by
\begin{equation}
                          f(\varepsilon)=\varepsilon/\zeta(\theta).
\label{Eq:s1-finite}
\end{equation}
This is illustrated in figure \ref{Fig:asymptotic-behaviour-parameter}.

\subsection{Infinite ACIM}\label{theta<1}
\begin{figure}[htb]
\begin{center}
\includegraphics[width=7cm]{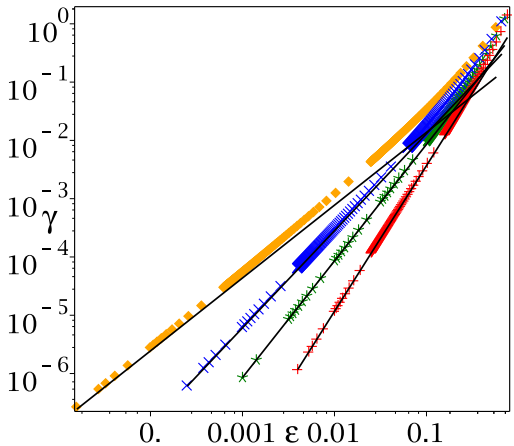}
\caption{Escape rate scaling with hole length. In this figure the numerically computed escape rate $\gamma$ is plotted as a function of hole size $\varepsilon$ for four values of the map parameter $\theta$; $0.4$ (red crosses), $0.5$ (green asterixs), $0.6$ (blue diagonal crosses) and $0.8$ (yellow diamonds). Also illustrated is the scaling and asymptotic behaviour with $\varepsilon$ as given by Eq.(\ref{Eq:f-infinite}) (black solid lines).  }\label{Fig:asymptotic-infinite}\end{center}
\end{figure}

When considering the case $\theta <1$ we can use the Euler-Maclaurin formula to rewrite the sum $ \sum_{k=1}^n \frac{1}{k^{\theta}}$ as
\begin{equation}
                          \sum_{k=1}^n\frac{1}{k^\theta}= \frac{n^{1-\theta}}{1-\theta}+\frac{n^{-\theta}}{2} -\frac{\theta }{12}n^{-\theta-1}+\frac{(\theta+2)(\theta+3)}{12(\theta-1)}+\frac{\theta(\theta+1)}{2} \int_1^n x^{-\theta-2}(\{x\}^2 -\{x\}+\frac{1}{6})dx.
\label{Eq:SumGrowth}
\end{equation}
The integral in Eq.(\ref{Eq:SumGrowth}) has the following bound
\begin{equation}
                          \frac{\theta(\theta+1)}{2} \int_1^n x^{-\theta-2}(\{x\}^2 -\{x\}+\frac{1}{6})dx \leq  \frac{\theta(\theta+1)}{12} \left(  \frac{-n^{-\theta -1}}{\theta+1} +\frac{1}{\theta+1}\right),
\label{Eq:Bound}
\end{equation}
which (using $\varepsilon=1/n ^{\theta}$)  implies that for $\theta <1$ we have
\begin{equation}
                          \sum_{k=1}^n\frac{1}{k^\theta}= \frac{n^{1-\theta}}{1-\theta}+ o\left(n^{1-\theta}\right).
\label{Eq:SumGrowth2}
\end{equation}
Considering then the $j=1$ term in Eq.(\ref{Eq:PArSum}) we let $f(n)=C/n$, with $C$ a constant. We then use
\begin{equation}
                         \sum_{k=j}^n{k-1 \choose j-1}\frac{1}{k^{\theta}}= \sum_{k=j}^n\frac{k^{j-1-\theta}}{(j-1)!}+o(k^{j-1-\theta})=\frac{n^{j-\theta}}{(j-1)!(j-\theta)}+o(n^{j-\theta}),
\label{Eq:Insumident}
\end{equation}
and
\begin{equation}
                          {n \choose j}=\frac{n^j}{j!}+o(n^j),
\label{Eq:binomIdent}
\end{equation}
in Eq.(\ref{Eq:PArSum}), whilst letting $n$ go to infinity.  which leaves us with
\begin{equation}
                         1-\theta\sum_{j=1}^{\infty}\frac{C^j}{j!(j-\theta)}=0 \label{Eq:ConstantFunc}
\end{equation}
In terms of $\varepsilon$ we have
\begin{equation}
                        f(\varepsilon)=C\varepsilon^{\frac{1}{\theta}},
\label{Eq:f-infinite}
\end{equation}
with $C$ a solution of Eq.(\ref{Eq:ConstantFunc}). From which we see that the escape rate scales polynomially with the hole length $\varepsilon$.
See figure \ref{Fig:asymptotic-infinite} for an illustration of this.
\section{Conclusion}
\label{Sec:Conclusion}
Here we investigated the small hole scaling of the escape rate, in a class of piecewise linear dynamical systems where the hole is placed in the neighbourhood of an indifferent fixed point, so as the hole shrinks the system converges to a closed system with intermittent behaviour. In some cases the intermittency can lead to the closed system preserving an infinite absolutely continuous invariant measure. By the introduction of a parameter dependence of the intermittency, we were able to control this particular property. In the finite measure preserving case we saw that the escape rate scaled linearly with the hole size and derived the asymptotic behaviour explicitly. In the infinite measure preserving case we saw that the escape rate scaled polynomially with the hole length whilst on the border between the two (where the measure is sometimes called ``barely infinite''), we saw logarithmic corrections to the scaling.
We saw that there is a non-trivial relationship between the properties of the absolutely continuous invariant measure of the closed system and the scaling of the escape rate given a suitably positioned hole in phase space. We would like to extend these results to more general systems, potentially allowing the study of the properties of infinite invariant measures via tools based upon escape rate theory and vice-versa.

A possible direction for further study in this direction is the following sequence of observations. One way to answer the question of  ``how infinite'' is a given system $(X, T, \mu)$, is to consider the wandering rate for some ``good'' measurable set $E$, which is defined to be the sequence $(w_n(E))_{n\geq1}$, where $w_n(E):=\mu\left( \bigcup_{k=0}^{n-1}T^{-k}(E)\right)$. In our case, we let the set $E$ be $A_1$, the first partition interval. Then, if $T=F_\alpha$, it is an easy calculation to see that the wandering rate is given by $w_n(A_1) = \sum_{k=1}^nt_k$. So, we have that for the $\alpha_H$-Farey map, the wandering rate is asymptotic to $\log n$, and for the examples from Section \ref{theta<1}, we have (either by comparing the sum to the integral or by considering the more careful calculation done above) that the wandering rate is asymptotic to $n^{1-\theta}$. One of the foundational results in infinite ergodic theory is Aaronson's ergodic theorem \cite{Aar97}, which states that there is no equivalent to Birkhoff's ergodic theorem, that is, that for infinite systems the pointwise behaviour of the ergodic averages is so complicated that it is impossible to find a normalising sequence to estimate their actual size. On the other hand, the same author showed that in the situation of a system with a wandering rate for a ``very good'' set that is regularly varying, there exists a sequence $(v_n)_{n\geq1}$ with the property that the ergodic averages rescaled using $v_n$ do have an asymptotic distribution. Slightly more precisely, if $\theta\in[0,1]$ and if $w_n(E)$ is regularly varying with exponent ${1-\theta}$, then the scaling sequence $v_n(E)$ is regularly varying with exponent $\theta$. Notice that for our examples in Section \ref{theta<1} we have that the scaling of the escape rate is the asymptotic inverse of this, i.e., it is equal to $1/\theta$. We tentatively conjecture that this relation might hold more generally.
\section*{Acknowledgments}
\label{Sec:Acknowledgements}
The authors would like to thank Giampaolo Cristadoro at the University of Bologna, Carl Dettmann at the University of Bristol and Alan Haynes at the University of York for enlightening conversations. GK would like to thank the University of Bristol for their hospitality during his visit there, similarly SM extends her thanks to the University of Bologna. The authors would also like to thank the referees for their careful reading and helpful comments which greatly improved the presentation.   

\bibliographystyle{unsrt}
\bibliography{Farey_Bib}

\end{document}